\documentclass[prd,twocolumn,showpacs,floatfix,amsmath,amssymb,floatfix]{revtex4}
\usepackage{graphicx,color,dcolumn,booktabs,bm}
\usepackage{longtable,lscape}
\usepackage{txfonts}
\usepackage{overpic}
\usepackage{amssymb}
\usepackage{indentfirst}
\usepackage{feynmf}   
\usepackage{slashed}  
\usepackage{cases}
\usepackage{color}
\usepackage{multirow}
\usepackage{epstopdf}
\usepackage{graphicx,color,dcolumn,booktabs,bm}

\graphicspath{{Figures/}} %
\usepackage[colorlinks,
                      citecolor=blue,
                      anchorcolor=red,
                      menucolor=red,
                      linkcolor=red,
                      filecolor=red,
                      runcolor=red,
                      urlcolor=blue,
                      frenchlinks=red]{hyperref}
\begin{document}
\def\pca{P_c^+(4380)}
\def\pcb{P_c^+(4450)}
\def\pcc{P_c^+ (4312)}
\def\pcd{P_c^+ (4440)}
\def\pce{P_c^+(4457)}

\title{Production of  $P_c $ states from $\Lambda_b$ decay}

\author{Qi Wu}
\author{Dian-Yong Chen} \email{chendy@seu.edu.cn}   
\affiliation{
 School of Physics, Southeast University,  Nanjing 210094, China\\}

\begin{abstract}
In the present work, we investigate $P_c(4312)$, $P_c(4440)$ and $P_c(4457)$  production from $\Lambda_b$ decay in a molecular scenario by using an effective Lagrangian approach.
We predict the ratio of the branching fraction of $\Lambda_b \to P_c K$, which is weakly dependent on our model parameter. We also find the ratios of the productions of the branching fractions of $\Lambda_b \to P_c K$ and $P_c \to J/\psi p$ can be well interpreted in the molecular scenario. Moreover, the estimated branching fractions of $\Lambda_b \to P_c K$ are of order $10^{-6}$, which could be tested by further measurements in LHCb Collaboration.

\end{abstract}


\maketitle

\section{Introduction}
\label{sec:introduction}

Searching for hadrons beyond 3-quark baryons and quark-antiquark mesons is one of intriguing frontier of hadron physics, even since the initial period of the quark model. Tremendous process has achieved in the recent decade. A growing number of tetraquark and pentaquark candidates have been observed experimentally (more details can be found in the recent review \cite{Liu:2019zoy, Chen:2016qju, Guo:2017jvc}). In 2015, the LHCb Collaboration reported two pentaquark candidates, $P_c(4380)$ and $P_c(4450)$, in the$J/\psi p $ invariant mass spectroscopy of $\Lambda_b \to K J/\psi p$ process \cite{Aaij:2015tga}. The two-body mass spectroscopy and angular distributions of three-body final states had been analyzed and the $J^P$ quantum numbers of these two tetraquark candidates are preferred to be of opposite parity has $J=3/2$ for one state and $J={5/2}$ for the other one.

It should be mentioned that before the LHCb observation,  there are theoretical predictions of molecular states composed by  a anti-charmed meson and charmed baryon \cite{Wu:2010jy, Yang:2011wz, Wang:2011rga, Wu:2012md}, such as $\Sigma_c \bar{D}$ and $\Lambda_c \bar{D}$, which may correspond to the observed $P_c$ states.  Further more, in the vicinity of $\pca$ and $\pcb$,  there are abundant thresholds of a baryon and a meson, such as $\Sigma_c^{(\ast)} \bar{D}^{(\ast)}$, $\Lambda_c \bar{D}^\ast$, $\chi_{c1} p$, $\psi(2S) p$. Along the way of molecular scenario, some interpretations related the above thresholds have been proposed \cite{Guo:2015umn,Chen:2015moa, Azizi:2016dhy, Roca:2015dva, Yang:2015bmv, Huang:2015uda, Chen:2016heh, Chen:2015loa, He:2015cea, Karliner:2015ina, Mironov:2015ica, Meissner:2015mza, Burns:2015dwa, Shimizu:2016rrd, Yamaguchi:2016ote, Eides:2015dtr}. It should be noticed that the masses of $\pca$ and $\pcb$ are above the threshold of $J/\psi$ and were observed in the $J/\psi p $ mode,  thus these two $P_c$ states more likely contain five constitute quarks, which is $c\bar{c} qqq$, where $q$ is up or down quark.  In the tetraquark scenario, a series of interpretations with different quark configurations to $\pca$ and $\pcb$ were proposed \cite{Maiani:2015vwa, Anisovich:2015cia, Li:2015gta, Ghosh:2015ksa, Wang:2015epa, Anisovich:2015zqa, Lebed:2015tna, Zhu:2015bba}.

\begin{table}[t]
 \centering
 \caption{The resonance parameters of the newly reported pentaquark states and the production ratio. \label{Tab:Para} }
 \begin{tabular}{p{1.5cm}<\centering p{2.2cm}<\centering p{2.2cm}<\centering p{2.2cm}<\centering}
 \toprule[1pt]
 State & Mass (MeV) & Width (MeV) & R($\%$) \\
 \midrule[1pt]
 $\pcc$ & $4311.9 \pm 0.7^{+6.8}_{-0.6}$  & $9.8 \pm 2.7 ^{+3.7}_{-4.5}$ & $0.30 \pm 0.07^{+0.34}_{-0.09}$ \\
 $\pcd$ &$4440.3 \pm 1.3^{+4.1}_{-4.7}$  & $20.6 \pm 4.9 ^{+8.7}_{-10.1}$ & $1.11 \pm 0.33^{+0.22}_{-0.10}$ \\
 $\pce$ & $4457.3 \pm 0.6^{+4.1}_{-1.7}$  & $6.4 \pm 2.0 ^{+5.7}_{-1.9}$ & $0.53 \pm 0.16^{+0.15}_{-0.13}$ \\
 \bottomrule[1pt]
 \end{tabular}
 \end{table}

Very recently, the LHCb Collaboration updated their analysis of the $J/\psi p$ invariant mass spectroscopy of $\Lambda_b \to K J/\psi p $ and find three pentaquark states, which are $\pcc$, $\pcd$ and $\pce$ \cite{Aaij:2019vzc}.  After the new observation, some interpretation have been proposed immediately in the molecular \cite{Chen:2019bip, Chen:2019asm, He:2019ify, Liu:2019tjn, Zhang:2019xtu, Meng:2019ilv, Mutuk:2019snd, Huang:2019jlf, Shimizu:2019ptd, Guo:2019kdc, Xiao:2019aya, Guo:2019fdo, Cheng:2019obk,  Eides:2019tgv} and tetraquark scenarios  \cite{Ali:2019npk, Giannuzzi:2019esi, Wang:2019got, Weng:2019ynv, Zhu:2019iwm, Cao:2019kst,Wang:2019krd}.  As listed in Table \ref{Tab:Para}, the mass of $\pcc$ is very close to the threshold of $\Sigma_c \bar{D}$, while $\pcd$ and $\pce$ are close to $\Sigma_c \bar{D}^\ast$ threshold, and the small mass splitting of $\pcd$ and $\pce$ may resulted from the spin-spin interactions of the components. Thus, one can assign $\pcc$ as $\Sigma_c \bar{D}$ molecular state with $J^P=\frac{1}{2}^-$, while $\pcd$ and $\pce$  are $\Sigma_c \bar{D}^\ast$ molecular states with $J^P=\frac{1}{2}^-$ and $\frac{3}{2}^-$, respectively. Such assignments are supported by the estimations in Refs. \cite{He:2019ify, Liu:2019tjn}.

Besides, the resonance parameters of the $P_c$ states, the production ratio, $R \equiv \mathcal{B} (\Lambda_c \to P_c K) \times \mathcal{B}(P_c \to J/\psi p)/\mathcal{B}(\Lambda_c \to J/\psi p K)$, were also measured, which are also listed in Table \ref{Tab:Para}. The new analysis indicates the production ratios are of order of one percent. The newly measured product ratios are much smaller than those for $\pca$ and $\pcb$ from their previous analysis, which are $(8.4 \pm 0.7 \pm 4.2) \%$ and $(4.1 \pm 0.5 \pm 1.1)\%$ for $\pca$ and $\pcb$, respectively. With the PDG average of the branching ratio $\mathcal{B}(\Lambda_b \to J/\psi p K)= (3.2^{+0.6}_{-0.5}) \times 10^{-4}$,  the production of the branching ratios for $\Lambda_b \to P_c K$ and $P_c \to J/\psi p$ are estimated to be,

\begin{widetext}
\begin{eqnarray}
\mathcal{B}(\Lambda_b \to \pcc K) \times \mathcal{B}(\pcc \to J/\psi p) &=& (0.96^{+1.12}_{-0.28}) \times 10^{-6},\nonumber\\
\mathcal{B}(\Lambda_b \to \pcd K) \times \mathcal{B}(\pcd \to J/\psi p) &=& (3.55^{+1.43}_{-1.20}) \times 10^{-6},\nonumber\\
\mathcal{B}(\Lambda_b \to \pce K)\times \mathcal{B}(\pce \to J/\psi p) &=& (1.70^{+0.77}_{-0.60} ) \times 10^{-6}.  \label{Eq:PR}
\end{eqnarray}
\end{widetext}
Besides the mass spectra of the $P_c$ states, how to understand the measured production ratios is an intriguing problem, which could help us to reveal the inner structures of the pentaquark states.  In Ref. \cite{Xiao:2019mvs}, the partial widths of  $P_c \to J/\psi p$ were estimated in a molecular scenario, thus, study the production process $\Lambda_b\to P_c K$ in the same molecular scenario and compared with the the measured production ratios listed in Eq. (\ref{Eq:PR}) can further test the molecular interpretations of $P_c$ states, which is the main task of the present work.

The present work is organized as follows. After introduction, the formula of the productions of  $\Lambda_b \to P_c K$ are present, including the related effective Lagrangians and production amplitudes. In section \ref{Sec:Num}, we present our numerical results and some discussions of the present results. A short summary is presented in Section \ref{Sec:Summary}.

\begin{figure}[htb]
\begin{tabular}{ccc}
  \centering
 \includegraphics[width=2.8cm]{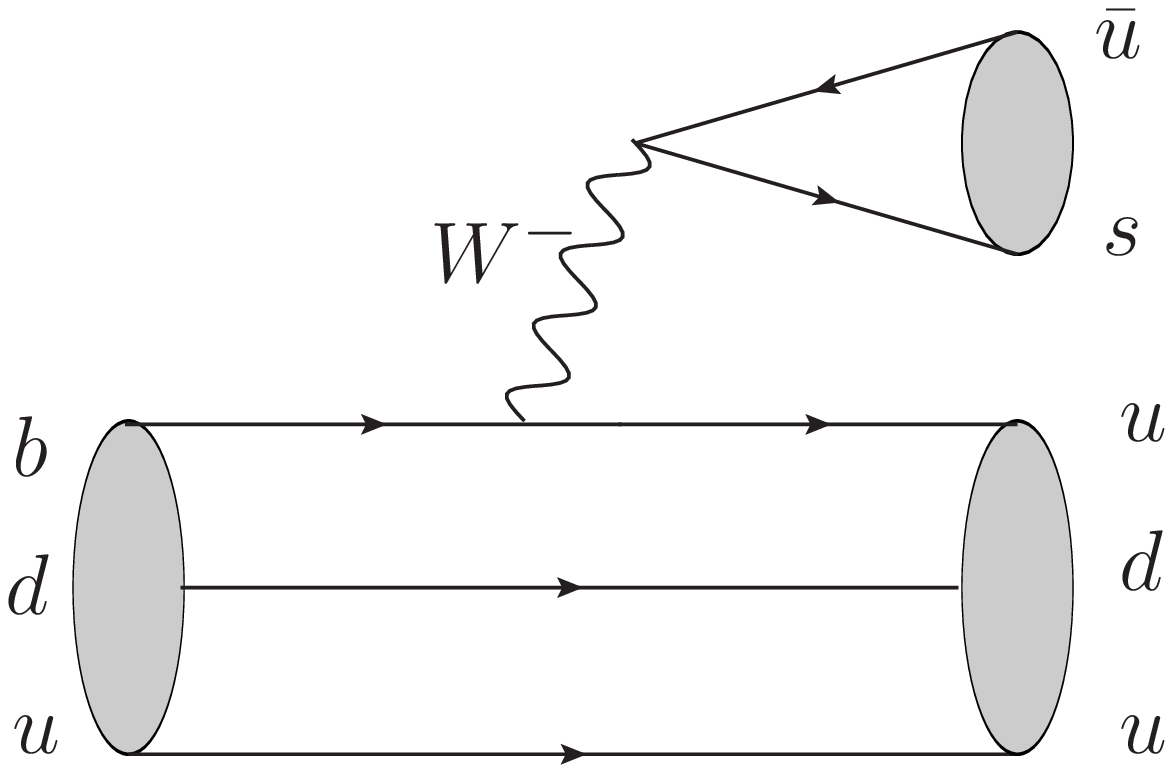}&
 \includegraphics[width=2.8cm]{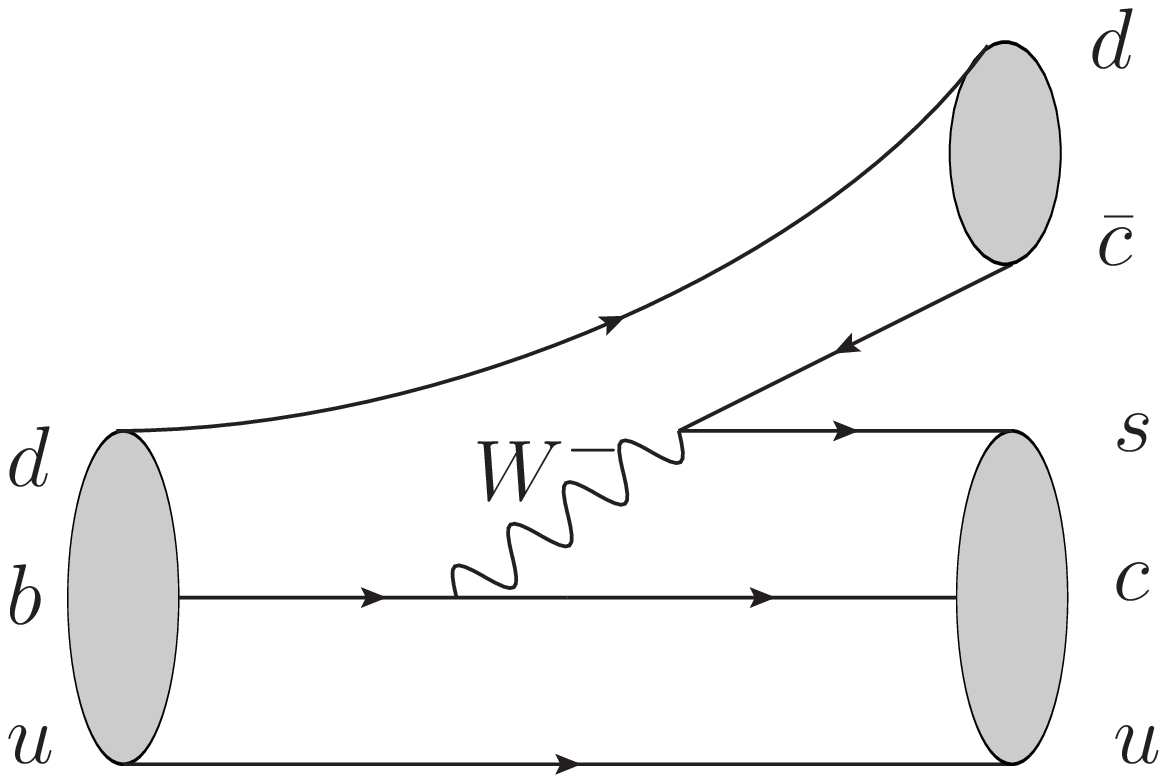}&
 \includegraphics[width=2.8cm]{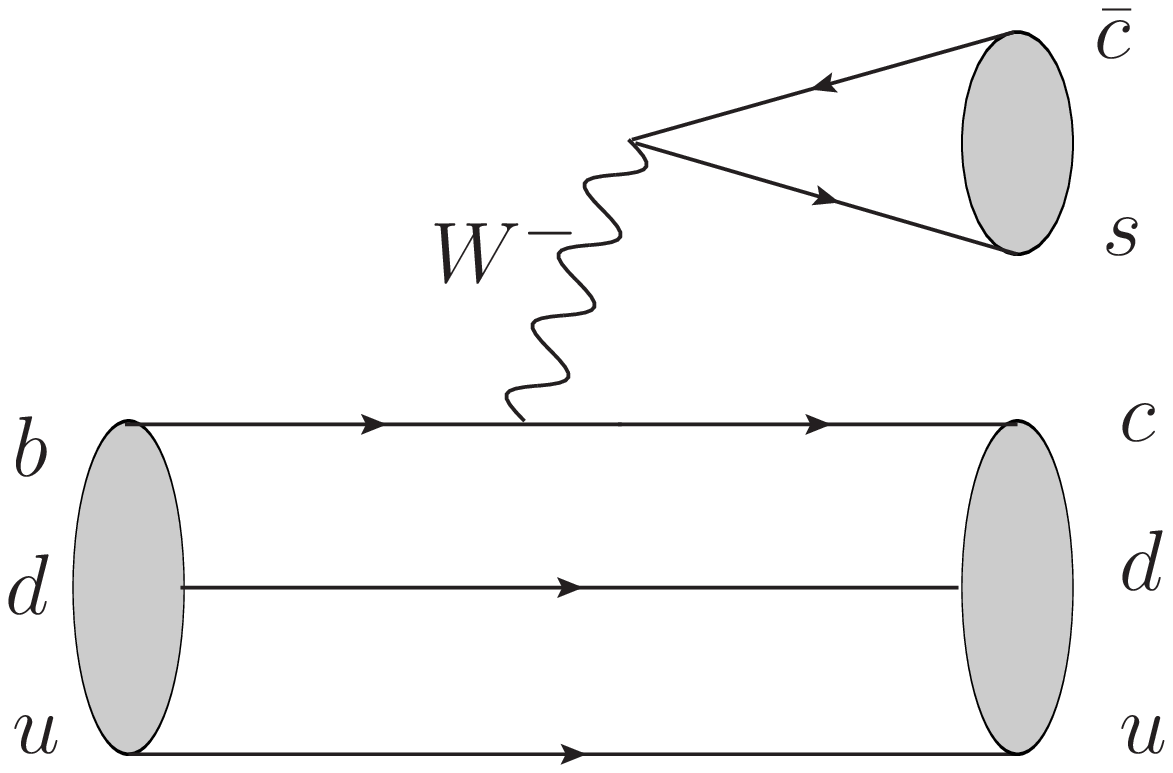}\\
 \\
 $(a)$ & $(b)$ &$(c)$
 \end{tabular}
 \caption{Possible diagrams contributing to  $\Lambda_b \to P_c K$ in quark level. \label{Fig:QL}}
\end{figure}

\section{The productions of $\Lambda_b \to P_c K$}
\label{Sec:Form}
We can first analyze the production process of $P_c$ states from the quark level. One should notice that $P_c$ states are produced accompany with a $K^-$ meson. In Fig. \ref{Fig:QL}-(a), the kaon is produced directly from $W^-$ meson. Since the $P_c$ states have a $c\bar{c}$ components, thus, the $b$ quark should transits to $u$ quark via $W^-$ emission, and the $c\bar{c}$ components are created from the vacuum. This kind of digram will be suppressed in the $P_c$ production, since $V_{ub}$ is about one order of magnitude smaller than $V_{cb}$.  In the second kind of mechanism as shown in Fig. \ref{Fig:QL}-(b),  the subprocess of the weak decay is $b \to c \bar{c} s$. The $\bar{c}$ quark and the $d$ quark in the initial $\Lambda_b$  form a anti-charmed meson, such as $\bar{D}^{(\ast)}$. The $cs$ quarks and the $u$ quark in the initial $\Lambda_b$ become a baryon, like $\Xi_c^{(\ast)}$. Then the $\Xi_c^{(\ast)}$ state emits a kaon and transits into $\Sigma_c$ and the recoiled $\Sigma_c$ and $\bar{D}^{(\ast)}$ form a $P_c$ state. In Fig. \ref{Fig:QL}-(c), the subprocess are the same as the one in Fig. \ref{Fig:QL}-(b), but the $\bar{c}s$ quark form a $\bar{D}_s^{(\ast)}$ and $cdu$ form a $\Sigma_c$. By emitting a kaon, $\bar{D}_s^{(\ast)-}$ meson transits into $\bar{D}^{(\ast)}$ and the recoiled $\bar{D}^{(\ast)}$ meson and $\Sigma_c$ form a $P_c$ state. Comparing to Fig. \ref{Fig:QL}-(c), the mechanism in Fig. \ref{Fig:QL}-(b) is suppressed due to color suppression in the hadronization process, thus in $\Lambda_b \to P_c K$ process, the mechanism in Fig. \ref{Fig:QL}-(c) is supposed to be dominant. In the present work, we estimate the process $\Lambda_b \to K P_c$ in the hadronic level and the related diagrams are listed in Fig. \ref{Fig:Tri}.

\subsection{Effective Lagrangians}

\begin{figure}[htb]
\begin{tabular}{ccc}
  \centering
 \includegraphics[width=2.9cm]{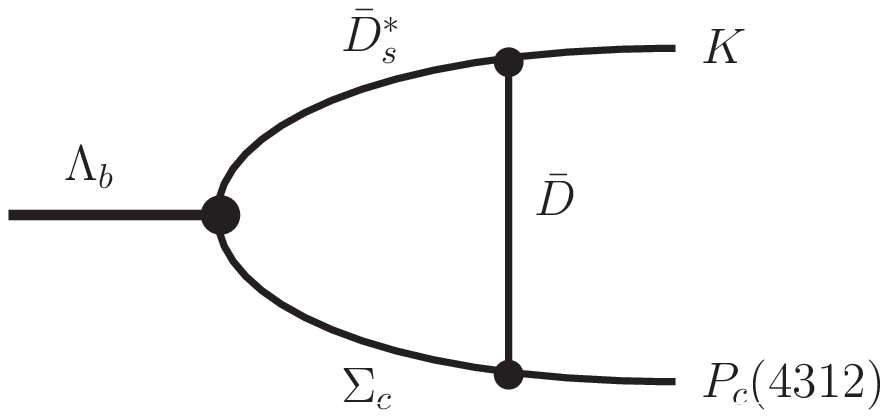}&
 \includegraphics[width=2.9cm]{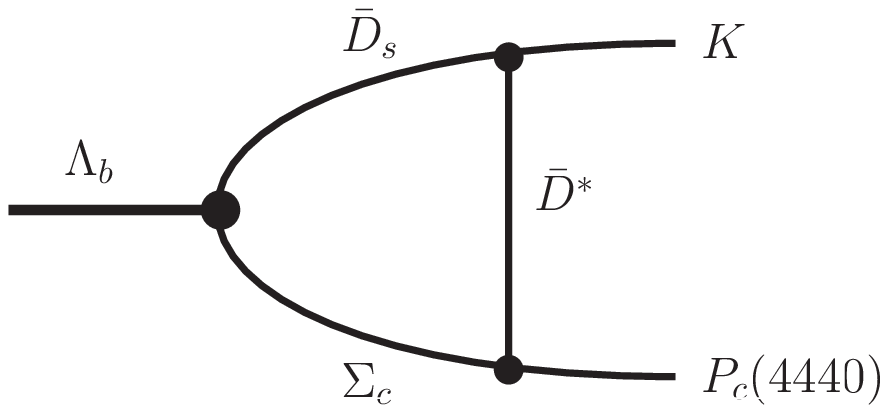}&
 \includegraphics[width=2.9cm]{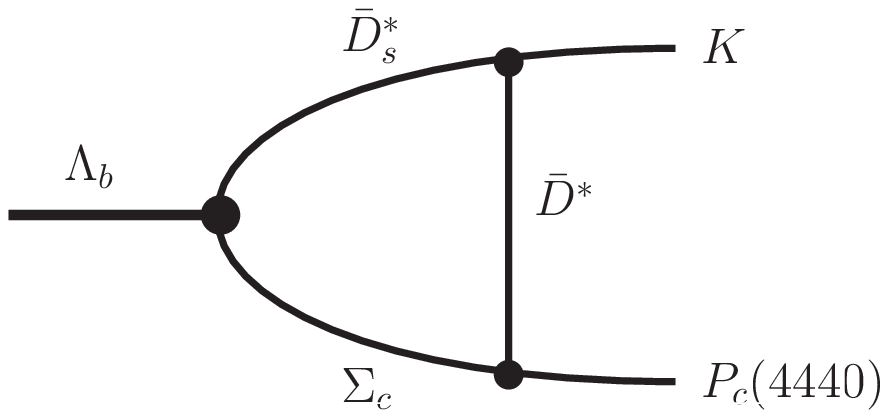}\\
 \\
 $(a)$ & $(b)$ &$(c)$\\
 \\
 \includegraphics[width=2.9cm]{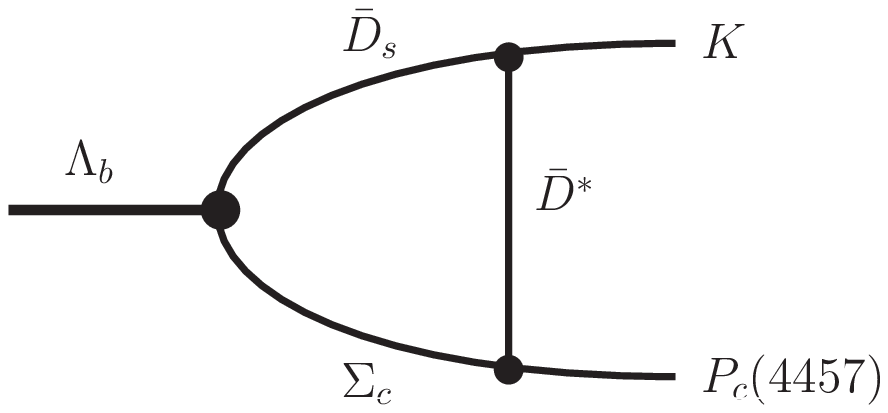}&
 \includegraphics[width=2.9cm]{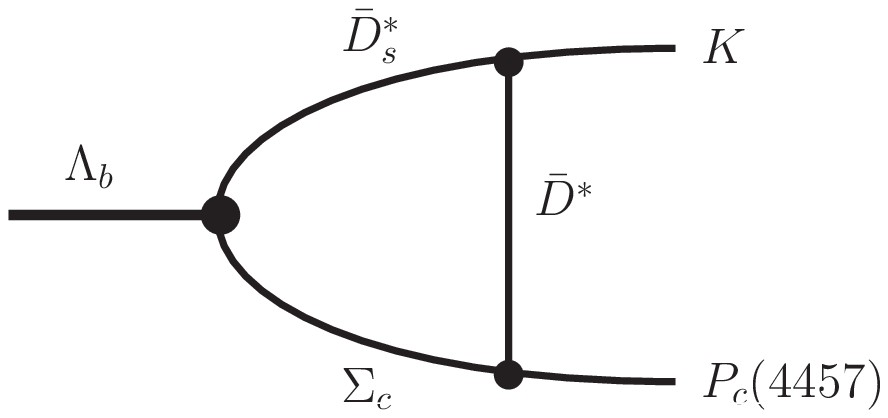}&\\
 \\
 $(d)$ & $(e)$&\\
 \end{tabular}
  \caption{Diagrams  contributing to $\Lambda_b\rightarrow P_c(4312) K$ (diagram(a)) $\Lambda_b\rightarrow P_c(4440) K$ (diagrams(b)-(c)) and $\Lambda_b\rightarrow P_c(4457) K$ (diagrams(d)-(e)).}\label{Fig:Tri}
\end{figure}

We employ an effective Lagrangian approach to estimate the diagrams in Fig. \ref{Fig:Tri}.
As for the $\Lambda_b \to \bar{D}_s^{(\ast)} \Sigma_c$,  the interactions vertexes are the same as the those of $\Lambda_b \to \bar{D}_s^{(\ast)} \Lambda_c$ and in the form \cite{Wang:2017mqp,Shi:2019hbf}
\begin{eqnarray}
\mathcal{L}_{\Lambda_b \Sigma_c D^\ast_s} &=& D^{\ast\mu}_s \bar{\Sigma}_c (A_1 \gamma_\mu \gamma_5+A_2\frac{p_{2\mu}}{m}\gamma_5+B_1 \gamma_\mu+B_2\frac{p_{2\mu}}{m})\Lambda_b,\nonumber\\
\mathcal{L}_{\Lambda_b \Sigma_c D_s} &=& i\bar{\Sigma}_c (A+B\gamma_5)\Lambda_b D_s,\label{eq:1}
\end{eqnarray}
where $A, \ B$, $A_1,\ A_2,\ B_1$ and $B_2$ are the recombinations of the form factors, which are,
\begin{eqnarray}
  A &=& -\lambda f_{D_s}[(m-m_2)f_1^{V}+\frac{m_1^2}{m}f_3^{V}],\nonumber\\
  B &=&-\lambda f_{D_s}[(m+m_2)f_1^{A}-\frac{m_1^2}{m}f_3^{A}],\nonumber\\
  A_1 &=& -\lambda f_{D^\ast_s}m_1[f_1^A+f_2^A\frac{m-m_2}{m}],\nonumber\\
  A_2&=& -2\lambda f_{D^\ast_s}m_1 f_2^A,\nonumber\\
  B_1 &=& \lambda f_{D^\ast_s}m_1[f_1^V-f_2^V\frac{m+m_2}{m}],\nonumber\\
   B_2&=& 2\lambda f_{D^\ast_s}m_1 f_2^V,
\end{eqnarray}
where $\lambda=\frac{G_F}{\sqrt{2}}V_{cb}V_{cs}a_1$. $m$, $m_1$ and $m_2$ is the mass of $\Lambda_b$, $D^{(\ast)}_s$ and $\Sigma_c$, respectively. $f_i^{(A,V)}$ (i=1,2,3) are the transition form factors of $\Lambda_b \to \Sigma_b$, which will be discussed in the next section.

The effective Lagrangians related to $D^{(\ast)}_s D^{(\ast)}K$ are \cite{Azevedo:2003qh},
\begin{eqnarray}
\mathcal{L}_{KDD^{\ast}_s}&=&i g_{KDD^{\ast}_s}D^{\ast\mu}_s[\bar{D}\partial_\mu\bar{K}-(\partial_\mu\bar{D})\bar{K}]+H.c.,\nonumber\\
\mathcal{L}_{KD_sD^{\ast}}&=&i g_{KD_sD^\ast}D^{\ast\mu}[\bar{D}_s\partial_\mu K -(\partial_\mu\bar{D}_s)K]+H.c.,\nonumber\\
\mathcal{L}_{KD^\ast_s D^{\ast}}&=&-g_{KD^\ast_s D^\ast}\epsilon^{\mu\nu\alpha\beta}(\partial_\mu\bar{D}^\ast_\nu \partial_\alpha D^\ast_{s\beta}\bar{K}+\partial_\mu D^\ast_{\nu}\partial_\alpha\bar{D}^\ast_{s\beta}K), \nonumber\\\label{eq:2}
\end{eqnarray}
where the coupling constant are $g_{KDD^{\ast}_s}=g_{KD_sD^\ast}=5.0$, $g_{KD^\ast_s D^\ast}=7.0\ \mathrm{GeV}^{-1}$. The effective Lagrangian of $P_c$ and $\Sigma_c D^{(\ast)}$ are \cite{Zou:2002yy}
\begin{eqnarray}
\mathcal{L}_{p_{c1}\Sigma_c D} &=& g_{p_{c1}\Sigma_c D}\bar{\Sigma}_c P_{1c} D^0 +H.c., \nonumber\\
\mathcal{L}_{p_{c2}\Sigma_c D^\ast} &=& g_{p_{c2}\Sigma_c D^\ast}\bar{\Sigma}_c \gamma_5 (g_{\mu\nu}-\frac{p_{4\mu}p_{4\nu}}{m^2_4})\gamma^\nu P_{c2} D^{\ast\mu},\nonumber\\
\mathcal{L}_{p_{c3}\Sigma_c D^\ast} &=& g_{p_{c3}\Sigma_c D^\ast}\bar{\Sigma}_c P_{c3\mu} D^{\ast\mu}.\label{eq:3}
\end{eqnarray}
where $p_{c1}$, $p_{c2}$ and $p_{c3}$ denotes $P_c(4312)$, $P_c(4440)$ and $P_c(4457)$ hereafter, respectively.

\subsection{Decay amplitudes}

With the effective Lagrangians listed above, we can obtain the amplitudes involve in the present work.  The decay amplitude of $\Lambda_b(p)\rightarrow D^\ast_s(p_1)\Sigma_c(p_2)[D(q)]\rightarrow K(p_3)P_{c1}(p_4)$ corresponding to Fig. \ref{Fig:Tri}-(a) is
\begin{eqnarray}
\mathcal{M}_a&=&i^3 \int\frac{d^4 q}{(2\pi)^4}[-ig_{p_{c1}\Sigma_c D}\bar{u}(p_4)](p_2\!\!\!\!\!\slash+m_2)[(A_1 \gamma_\mu \gamma_5+A_2\frac{p_{2\mu}}{m}\gamma_5\nonumber\\
&&+B_1 \gamma_\mu+B_2\frac{p_{2\mu}}{m})u(p)][-g_{KDD^{\ast}_s}(p_3-q)_\nu](-g^{\mu\nu}+\frac{p^\mu_1 p^\nu_1}{m^2_1})\nonumber\\
&&\times\frac{1}{p^2_1-m^2_1}\frac{1}{p^2_2-m^2_2}\frac{1}{q^2-m^2_E}\mathcal{F}(q^2,m^2).
\end{eqnarray}

The decay amplitude of $\Lambda_b(p)\rightarrow D^{(\ast)}_s(p_1)\Sigma_c(p_2)[D^\ast(q)]\rightarrow K(p_3)P_{c2}(p_4)$ corresponding to Fig. \ref{Fig:Tri}-(b) and (c) are
\begin{eqnarray}
\mathcal{M}_b&=&i^3 \int\frac{d^4 q}{(2\pi)^4}[g_{p_{c2}\Sigma_c D^\ast}\bar{u}(p_4)\gamma^\nu \gamma_5(g_{\mu\nu}-\frac{p_{4\mu}p_{4\nu}}{m^2_4})](p_2\!\!\!\!\!\slash+m_2)\nonumber\\
&&\times[i(A+B\gamma_5)u(p)][-g_{KD^\ast D_s}(p_1+p_3)_\alpha](-g^{\mu\alpha}+\frac{q^\mu q^\alpha}{m^2_E}) \nonumber\\
&&\times\frac{1}{p^2_1-m^2_1}\frac{1}{p^2_2-m^2_2}\frac{1}{q^2-m^2_E}\mathcal{F}(q^2,m^2)\nonumber\\
\mathcal{M}_c&=&i^3 \int\frac{d^4 q}{(2\pi)^4}[g_{p_{c2}\Sigma_c D^\ast}\bar{u}(p_4)\gamma^\nu \gamma_5(g_{\mu\nu}-\frac{p_{4\mu}p_{4\nu}}{m^2_4})](p_2\!\!\!\!\!\slash+m_2)\nonumber\\
&&\times[(A_1 \gamma_\alpha \gamma_5+A_2\frac{p_{2\alpha}}{m}\gamma_5+B_1 \gamma_\alpha+B_2\frac{p_{2\alpha}}{m})u(p)]\nonumber\\
&&\times[-g_{KD^\ast D^\ast_s}\varepsilon_{\rho\lambda\eta\tau}q^\rho p^\eta_1](-g^{\mu\lambda}+\frac{q^{\mu}q^{\lambda}}{m^2_E})(-g^{\alpha\tau}+\frac{p_1^{\alpha}p_1^{\tau}}{m^2_1})\nonumber\\
&&\times\frac{1}{p^2_1-m^2_1}\frac{1}{p^2_2-m^2_2}\frac{1}{q^2-m^2_E}\mathcal{F}(q^2,m^2).
\end{eqnarray}

The decay amplitudes of $\Lambda_b(p)\rightarrow D^{(\ast)}_s(p_1)\Sigma_c(p_2)[D^\ast(q)]\rightarrow K(p_3)P_{c3}(p_4)$ corresponding to Fig. \ref{Fig:Tri}-(d) and (e) are
\begin{eqnarray}
\mathcal{M}_d&=&i^3 \int\frac{d^4 q}{(2\pi)^4}[-ig_{p_{c3}\Sigma_c D^\ast}\bar{u}_\mu(p_4)](p_2\!\!\!\!\!\slash+m_2)[i(A+B\gamma_5)\nonumber\\
&&u(p)][-g_{KD^\ast D_s}(p_1+p_3)_\nu](-g^{\mu\nu}+\frac{q^{\mu}q^{\nu}}{m^2_E})\nonumber\\
&&\times\frac{1}{p^2_1-m^2_1}\frac{1}{p^2_2-m^2_2}\frac{1}{q^2-m^2_E}\mathcal{F}(q^2,m^2) \nonumber\\
\mathcal{M}_e&=&i^3 \int\frac{d^4 q}{(2\pi)^4}[-ig_{p_{c3}\Sigma_c D^\ast}\bar{u}_\sigma(p_4)](p_2\!\!\!\!\!\slash+m_2)\nonumber\\
&&\times[(A_1 \gamma_\rho \gamma_5+A_2\frac{p_{2\rho}}{m}\gamma_5+B_1 \gamma_\rho+B_2\frac{p_{2\rho}}{m})u(p)]\nonumber\\
&&\times[-g_{KD^\ast D^\ast_s}\varepsilon_{\mu\nu\alpha\beta}q^\mu p^\alpha_1](-g^{\sigma\nu}+\frac{q^\sigma q^\nu}{m^2_E})(-g^{\rho\beta}+\frac{p^\rho_1 p^\beta_1}{m^2_1})\nonumber\\
&&\times\frac{1}{p^2_1-m^2_1}\frac{1}{p^2_2-m^2_2}\frac{1}{q^2-m^2_E}\mathcal{F}(q^2,m^2).
\end{eqnarray}
In the present work, a monopole form factor is introduced to depict the off-shell effect of the exchanged mesons, which is,
\begin{eqnarray}
\mathcal{F}(q^2,m^2) =\frac{m^2 -\Lambda^2}{q^2-\Lambda^2},
\end{eqnarray}
where $\Lambda=m+\alpha \Lambda_{QCD}$, $\Lambda_{QCD}=220 \ \mathrm{MeV}$ and $\alpha$ is a model parameter, which is of order of unit \cite{Tornqvist:1993vu, Tornqvist:1993ng, Locher:1993cc, Li:1996yn}.

With above amplitudes, one can estimated the partial width of $\Lambda_b\rightarrow P_c K$ by
\begin{eqnarray}
\Gamma_{\Lambda_b} &=& \frac{1}{2}\frac{1}{8\pi} \frac{|\vec{p}|}{m^2}\overline{|\mathcal{M}|^2}
\end{eqnarray}
where the factor $1/2$ results from the average of $\Lambda_b$ spin and $\vec{p}$ is the momentum of $P_c$ or $K$ in the rest frame of $\Lambda_b$.  The overline indicates the sum over the spins of final states.

\section{Numerical Results and discussions}
\label{Sec:Num}

Before we estimate the partial width of $\Lambda_b \to P_c K$ in the present scenario, we first discuss the transition form factors of $\Lambda_b \to \Sigma_c$. Unfortunately, there are no direct estimation of these transition form factors. One should be notice, the constitute quarks and the spatial part of the $\Sigma_c$ and $\Lambda_c$ are the same, thus the transition form factors of $\Lambda_b \to \Sigma_c$ should be the same as those of $\Lambda_b \to \Lambda_c$, but smaller in magnitude due to light quark spin flipping in the transition $\Lambda_b\ \to \Sigma_c$.  Here we defined the suppress ratio $R$ as,
\begin{eqnarray}
f_i^{A,V}(Q^2)= R F_i^{A,V}(Q^2), \ \ \ \{i=1,2,3\}
\end{eqnarray}
where $f_i^{(A,V)}$ and $F_i^{(A,V)}$ with $i=(1,23)$ are the transition form factors of $\Lambda_b \to \Sigma_c$ and $\Lambda_b \to \Lambda_c$, respectively.  The details of transition form factors of $\Lambda_b\to \Lambda_c$ are presented in Appendix \ref{Sec:App1}. Furthermore, the coupling constants related to $P_c$ and $\Sigma_c \bar{D}^{(\ast)}$ will be discussed later.

\begin{figure}[htb]
  \centering
 \includegraphics[width=0.90\linewidth]{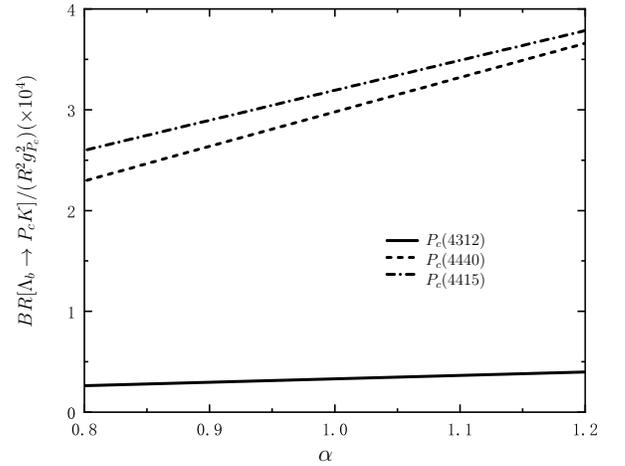}
  \caption{The $\alpha$-dependence of the branching fractions of  $\Lambda_b\rightarrow P_{c1} K$ (straight line),  $\Lambda_b\rightarrow P_{c2} K$ (dashed line) and $\Lambda_b\rightarrow P_{c3} K$ (dotted line).}\label{fig:ratios}
\end{figure}

In Fig~\ref{fig:ratios}, we plot the $\alpha$-dependence of $\mathcal{B}(\Lambda_b \to P_c K)/(R^2 g_{P_c}^2)$, which are of order   $10^{-4}$ for $\pcd$ and $\pce$ and  $10^{-5}$ for $\pcc$, respectively.  As for the coupling constants $g_{P_c}$, they could be estimated by the compositeness condition with the assumption that all three observed $P_c$ states are molecular states. In Ref. \cite{Xiao:2019mvs}, the coupling constants are estimated depending on a model parameter $\Lambda$, which is of order one GeV. When one take $\Lambda=1$ GeV, the coupling constants are estimated to be, $g_{P_{c1}} =2.25$, $g_{P_{c2}} =1.72$ and $g_{P_{c3}} =1.77$, respectively, which are very similar to those in Ref. \cite{Guo:2019kdc}.  With above coupling constants, the ratio of the branching fractions $\Lambda_b\to P_c K$ are estimated to be,
\begin{eqnarray}
\mathcal{R}^{\rm P}_{12}&\equiv& \frac{\mathcal{B}(\Lambda_b \to P_{c1} K)}{\mathcal{B}(\Lambda_b \to P_{c2} K)}= 0.19^{+0.01}_{-0.00},\nonumber\\
\mathcal{R}^{\rm P}_{13}&\equiv& \frac{\mathcal{B}(\Lambda_b \to P_{c1} K)}{\mathcal{B}(\Lambda_b \to P_{c3} K)}= 0.17^{+0.00}_{-0.01},\nonumber\\
\mathcal{R}^{\rm P}_{23}&\equiv& \frac{\mathcal{B}(\Lambda_b \to P_{c2} K)}{\mathcal{B}(\Lambda_b \to P_{c3} K)}= 0.88^{+0.03}_{-0.05},
\label{Eq:B12}
\end{eqnarray}
which are independent on $R$. The center values correspond to $\alpha=1.0$ and the uncertainties are resulted from the variation of model parameter $\alpha$ from 0.8  to 1.2. Our estimation indicates the production ratio are very weakly dependent on the model parameter.

\begin{figure}[htb]
  \centering
 \includegraphics[width=0.90\linewidth]{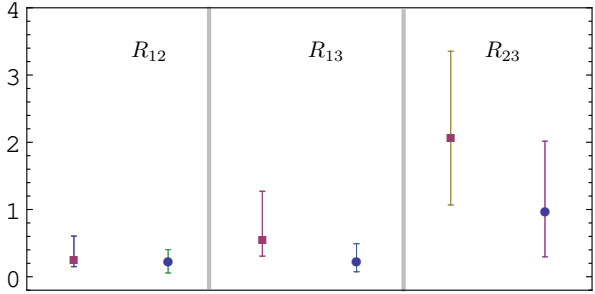}
  \caption{A Comparison of the ratio $R_{ij}$. The squares with error bar are measured data from LHCb Collaboration \cite{Aaij:2019vzc} and the full circles with error bar are the estimations in the present work. }\label{fig:com}
\end{figure}

Our estimation indicates that product ratios are very weakly dependent on the model parameter $\alpha$. In Ref. \cite{Xiao:2019mvs}, the partial widths of $P_c(4320) \to J/\psi p$, $P_c(4440) \to J/\psi p$ and $P_c(4450) \to J/\psi p$ are estimated to be  $5.6$, $9.3$ and $2.6$ MeV , respectively, when we take $\Lambda=1$ GeV. With these estimated partial widths and the measured total widths of $P_c$ states, one can get the branching fraction of $P_c \to J/\psi p$, which are,
 \begin{eqnarray}
 \mathcal{B}(P_c(4312) \to J/\psi p) &=& 0.57^{+0.27}_{-0.31}, \nonumber\\
  \mathcal{B}(P_c(4440) \to J/\psi p) &=& 0.45^{+0.22}_{-0.25}, \nonumber\\
 \mathcal{B}(P_c(4457) \to J/\psi p) &=& 0.41^{+0.38}_{-0.18},
\label{Eq:DBR}
 \end{eqnarray}
and then the decay ratio are,
\begin{eqnarray}
\mathcal{R}_{12}^{\rm D}= \frac{P_{c1} \to J/\psi P}{P_{c2} \to J/\psi P} &=&1.27^{+0.85}_{-0.96}, \nonumber\\
\mathcal{R}_{13}^{\rm D}= \frac{P_{c1} \to J/\psi P}{P_{c3} \to J/\psi P} &=& 1.41^{+1.48}_{-0.97}, \nonumber\\
\mathcal{R}_{23}^{\rm D}= \frac{P_{c2} \to J/\psi P}{P_{c3} \to J/\psi P} &=& 1.11^{+1.18}_{-0.77} .
\end{eqnarray}

With the production and decay ratios estimated in the molecular scenario, we can get the product of the product and decay ratios, i.e.,
\begin{eqnarray}
R_{ij}= \mathcal{R}_{ij}^{\rm P} \times \mathcal{R}_{ij}^{\rm D}
\end{eqnarray}
and compare these ratios with the experimental measurement as listed in Eq. (\ref{Eq:PR}). We present a comparison of the measured $R_{ij}$ from LHCb Collaboration \cite{Aaij:2019vzc} and the estimation in the present work in Fig. \ref{fig:com}. One can find our estimations are consistent with the experimental data from LHCb Collaboration within error, which indicates all three $P_c$ states could be interpreted as molecular states.

Moreover, taking the branching ratios estimated in the molecular scenario as listed in Eq. (\ref{Eq:DBR}) back to Eq. (\ref{Eq:PR}) one can get the branching ratios of $P_c $ production, which are,
\begin{eqnarray}
\mathcal{B}(\Lambda_b \to \pcc^+ K) &=&(1.68^{+2.11}_{-1.02}) \times 10^{-6}, \nonumber\\
\mathcal{B}(\Lambda_b \to \pcd^+ K) &=& (7.86^{+4.96}_{-5.04}) \times 10^{-6}, \nonumber\\
\mathcal{B}(\Lambda_b \to \pce^+ K) &=& (4.18^{+4.38}_{-2.33}) \times 10^{-6}.
\end{eqnarray}
 In the present scenario, the ratio $\mathcal{B}(\Lambda_b \to P_c K)/(R^2 g_{P_c}^2) $ are estimated to be of  $10^{-5}$ for $\pcc$ and $10^{-4}$ for $\pcd$ and $\pce$, respectively. Considering the light quark spin flip suppression in the $\Lambda_b \to \Sigma_c D^{(\ast)}$ process, one can suppose $R$ much smaller than one. In the present work, the estimated product ratio could be consistent with the experimental measurements within error when one take $R=0.09$, which are,
 \begin{eqnarray}
\mathcal{B}(\Lambda_b \to \pcc^+ K) &=&(1.11 \sim 1.67) \times 10^{-6}, \nonumber\\
\mathcal{B}(\Lambda_b \to \pcd^+ K) &=& (5.65\sim 9.02) \times 10^{-6}, \nonumber\\
\mathcal{B}(\Lambda_b \to \pce^+ K) &=& (6.78\sim 9.88) \times 10^{-6},
\end{eqnarray}
respectively.

It is interesting to note that the $\Lambda_b\to \Sigma_c$ transition requires the spin flip of the involved light quark system. This has been widely expected to be power suppressed in $1/m_b$. A relevant counterpart is the transition of bottom meson into a charmed scalar meson~\cite{Shen:2012mm}. An analysis in the QCD sum rules indicates that the helicity-flipped transition form factor is smaller than the ordinary heavy-to-light transition form factor by a factor 3 to 5.  A similar power suppression  might also happen in $\Lambda_b\to \Sigma_c$ compared to $\Lambda_b\to \Lambda_c$. If this were true, it indicates the use of $R\sim0.1$ is reasonable.

\section{Summary}
\label{Sec:Summary}
In the present work, we estimated the $P_c$ production from  $\Lambda_b$  decay in $P_c$ molecular scenario, where $\pcc$ is considered as $\Sigma_c \bar{D}$ molecular with $J^{P}=\frac{1}{2}^{-}$, while $\pcd$ and $\pce$ are interpreted  as  $\Sigma_c \bar{D}^\ast$ molecule with $J^P=\frac{1}{2}^-$ and $\frac{3}{2}^-$, respectively.  By analyzing the production process in quark level, we find the production process occur via the following process, $\Lambda_b$ could couple with  $\Sigma_c \bar{D}_s^{(\ast)}$  and the $\bar{D}_s^{(\ast)}$ transits into $\bar{D}^{(\ast)}$ via kaon emission and the recoil $\bar{D}^{(\ast)}$ and $\Sigma_c$ couple to $P_c$ state.

The $P_c$ production process are investigated in hadronic level with an effective Lagrangian approach. Unfortunately, the transition form factors related to $\Lambda_b \to \Sigma_c$ are unknown. In the present work, we borrow the form factors of $\Lambda_b \to \Lambda_c$ since the flavor and spatial parts of $\Lambda_c$ and $\Sigma_c$ are the same.  However, the transition form factors of $\Lambda_b \to \Sigma_c$ should be smaller than those of $\Lambda_b\to \Lambda_c$ since the suppression caused by the light quark spin flip. Here, we define a suppression factor $R$. Our estimation indicates the $R$ independent ratios of $\mathcal{B}(\Lambda_b \to P_c K) \times \mathcal{B}(P_c \to J/\psi p)$ in the molecular scenario are consistent with the experimental measurements from LHCb Collaboration \cite{Aaij:2019vzc}. Moreover, we find the magnitude of the production of the branching ratios for $\Lambda_b\to P_c K$ and $P_c \to J/
\psi p$ could be reproduced when one take the suppression factor  $R=0.09$.

In the molecular scenario, the branching ratio of $\Lambda_b\to P_c K$ are estimated. Together with the branching ratio of $P_c \to J/\psi p$ estimated in Ref. \cite{Xiao:2019mvs}, we find we can interpret the decay and production properties of $P_c$ states simultaneously in the molecular scenario, which indicates that $P_c$ states could be good candidates of $\Sigma_c \bar{D}^{(\ast)}$. Furthermore, in the present work and in Ref. \cite{Xiao:2019mvs}, we present our estimation of the production and decay branching ratios, which could be tested by further analysis in the LHCb Collaboration.

\section*{Acknowledgement}
\label{sec:acknowledgement}
The authors would like to thank Wei Wang for useful discussion. This work is supported in part by the National Natural Science Foundation of China (NSFC) under Grant Nos. 11775050\\

\appendix

\section{The transition form factors of $\Lambda_b\to \Lambda_c$ }
\label{Sec:App1}

The transition form factor pf $\Lambda_b\to \Lambda_c$ could be parameterized in the form \cite{Gutsche:2015mxa},
\begin{eqnarray}
F(Q^2)_i^{A,V}=\frac{F(0)}{1-a\zeta+b\zeta^2},
\label{Eq:A1}
\end{eqnarray}
where $\zeta=Q^2/m^2$. In Table. \ref{Tab:FFs}, we collect the parameters related to the transition form factors of $\Lambda_b\to \Lambda_c$ \cite{Gutsche:2015mxa}.

\begin{table}[htb]
\begin{center}
\caption{The values of the parameters $F(0)$, $a$ and $b$ in the form factors of $\Lambda_b\rightarrow\Lambda_c$ transition\cite{Gutsche:2015mxa}.}\label{Tab:FFs}
  \setlength{\tabcolsep}{2.4mm}{
\begin{tabular}{ccccccc}
  \toprule[1pt]
                  & $F^V_1$ & $F^V_2$ & $F^V_3$ & $F^A_1$ & $F^A_2$ & $F^A_3$ \\
  \midrule[1pt]
           $F(0)$ & 0.549 & 0.110 & -0.023 & 0.542 & 0.018 & -0.123 \\
           $a$    & 1.459 & 1.680 & 1.181  & 1.443 & 0.921 & 1.714  \\
           $b$    & 0.571 & 0.794 & 0.276  & 0.559 & 0.255 & 0.828  \\
  \bottomrule[1pt]
\end{tabular}}
\end{center}
\end{table}
where $F^V_i$ and $F^A_i$ (i=1,2,3) are the form factors of $\Lambda_b\rightarrow\Lambda_c$.

In the present estimation, we further parameterize the form factors in the form,
\begin{eqnarray}
F(Q^2)= F(0)\frac{\Lambda^2_1}{Q^2-\Lambda^2_1}\frac{\Lambda^2_2}{Q^2-\Lambda^2_2},\label{Eq:A3}
\end{eqnarray}
which can avoid ultraviolet divergence in the loop integrals and evaluate the loop integrals with Feynman parameterization methods. The values of $\Lambda_1$ and $\Lambda_2$ in above form factor are obtained by fitting Eq.~(\ref{Eq:A1}) with Eq.~(\ref{Eq:A3}) and the fitted parameter values are list in Table~\ref{Table:PARA2}.

\begin{table}[htb]
\begin{center}
\caption{Values of  parameters $\Lambda_1$ and $\Lambda_2$ in unit of GeV. }\label{Table:PARA2}
  \setlength{\tabcolsep}{2.4mm}{
\begin{tabular}{ccccccc}
 \toprule[1pt] 
  Parameter       &$F^{V/A}_1$ & $F^{V/A}_2$ & $F^{V/A}_3$\\
  \midrule[1pt]
  $\Lambda_1$     & 6.613/6.649  & 6.218/8.320 & 7.268/6.160\\   $\Lambda_2$     & 6.598/6.635  & 6.146/8.246 & 7.223/6.085\\
\bottomrule[1pt]
\end{tabular}}
\end{center}
\end{table}


\end{document}